\documentclass{article}
\newcommand{\bpi}{\mbox{\boldmath $\pi$}}

\def\bea{\begin{eqnarray}}
\def\eea{\end{eqnarray}}

\def\be{\begin{equation}}
\def\ee{\end{equation}}

\def\DR{\rm I\kern-1.45pt\rm R}
\begin{document}
\begin{center}
 {\Large\bf  External field influence on semiflexible macromolecules: geometric coupling}\\ [5mm]

 {\large Stefano Bellucci$\;^1$, Yevgeny Mamasakhlisov$\;^2$ and Armen Nersessian$\;^2$ }\\[5mm]

{\sl $\;^1$INFN-Laboratori Nazionali di Frascati, Via E. Fermi 40,
00044 Frascati, Italy}\\

 {\sl $\;^2$ Yerevan
State University, A.Manoogian St. 1, 0025, Yerevan, Armenia}
\end{center}
\begin{abstract}
We suggested a geometric approach to address the external field influence on the DNA
molecules, described by the WLC model via geometric coupling.
It consists in the introduction of the effective metrics depending on the potential of the external field,
with further re-definition of the arc-length parameter and of the  extrinsic curvatures
of the DNA molecules. It yields the nontrivial impact of the external field in the internal energy of macromolecules.
We give the Hamiltonian formulation of this model and perform its preliminary analysis
in the redefinition of the initial energy density.

\end{abstract}
\thispagestyle{empty}
\section{Introduction }
In recent years, single-molecule techniques have evolved into a
powerful toolset for studying the dynamical behavior of many
biological processes. It is now possible to follow on the nanometer
and millisecond scales the individual trajectory of a single enzyme
as it catalyzes a reaction, a molecular motor as it translocates, or
a single polypeptide or nucleic acid molecule as it unfolds and
refolds \cite{ritort,cornish,greenleaf}.

One of the widely used applications of single-molecule manipulations
is the direct investigation of double-stranded DNA (dsDNA). The
single-molecule stretching experiments gave a big contribution to
the understanding of the dsDNA structure and functionality
\cite{bauman+,10tension}. The interpretation of any stretching
experiment is closely related with the mechanism of elasticity. For
example, the dsDNA stretching experiments are usually interpreted in terms
of worm-like chains (WLC) \cite{bauman+,10tension,todd_rau,mamas},
while the ssDNA or ssRNA are usually described by freely jointed
chains with elastic bonds (EFJC) \cite{smith}.

The WLC  Hamiltonian is presented as a  functional on a differentiable
space curve $\gamma$ of fixed length $L$,
\begin{equation}\label{wlc}
{\cal E}_{0}^{WLC}=\frac{l_{p}k_{B}T}{2}\int_\gamma \kappa_{1}^2 ds\;,
\end{equation}
where  ${\bf r}(s)$ is the radius vector of an arbitrary point of
the curve as a function of the contour distance $s$ from one end to
that point, $T$ is the temperature, $k_{B}$ is the Boltzmann
constant, $l_{p}$ is the persistence length and $\kappa_1$ is the
first curvature (rigidity) of the curve, which is defined by the use of the
unit vector  ${\bf e}_1(s)$, ${\bf e}_{1}^2=1$
 tangential to the curve: $
\kappa_1=\vert{d{\bf e}_1}/{ds}|$, ${\bf e}_1=\frac{d{\bf r}}{ds}$,
where $s$ is the arc-length  of the curve ${\bf r}(\tau)$, \be
 s=\int_\gamma |d{\bf r}|, \qquad ds=|\frac {d\bf r}{d\tau}|d\tau
 \ee
An example of the spatial constraints, imposed on the
polymer chain, is the helical structure observed in many biological
and synthetic polymers \cite{gross}. In order to describe the helical
structure of the chain space, following Bugle and
Fujita \cite{bugl} (see also \cite{yamakawa1}), we need to introduce the unit curvature vector
${\bf e}_2=\frac{{d{\bf e}}_1/ds}{|d{\bf e}_1/ds|}$ and
define the potential energy of the chain  as follows:
\begin{equation}\label{wlc2}
{\cal E}^{BF}_0=\frac{k_{B}T}{2}\int_\gamma
ds\bigg[(\kappa_1-\alpha)^2+(\kappa_2-\beta)^2\bigg],
\end{equation}
where $\kappa_1={\dot {\bf e}}_1\cdot{\bf e}_2$ is a first
curvature (rigidity) of $\gamma$ and   $\kappa_2=\sqrt{|d{\bf
e}_2/ds|^2 -\kappa^2_1}$ is a second curvature (torsion) of the curve
$\gamma$. The $\alpha$ and $\beta$ parameters are supposed to be the
equilibrium bending and torsion constants. The absolute minimum of
the potential energy is given by $\kappa_1=\alpha$ and
$\kappa_2=\beta $ and corresponds to the regular helix with radius
$r_0=\frac{\alpha}{\alpha^2+\beta^2}$ and pitch
$h=\frac{2\pi|\beta|}{\alpha^2+\beta^2}$.

In this respect the paper by Feoli, Nesterenko and Scarpetta \cite{nesterenko} deserves
to be mentioned. There the authors suggested to describe the DNA
molecules by the functional
\be\label{feoli}
{\cal E}_0=\int_\gamma(c_0+c_1\kappa_1) ds,
\ee
 whose  minimum  corresponds to the helical configuration of
the curve. Unfortunately, these authors interpreted the functional
(\ref{gactions}) as a free energy of the DNA molecules, which yields
some  inconsistencies in the interpretation of the results.
 The importance  of that paper is not only in the
suggestion to describe the DNA molecules by the functionals linear
on the curvature, but to operate with the functionals depending on extrinsic curvatures by the use of the Dirac's theory of
constrained systems \cite{dirac}. Further below, we will discuss this issue in more detail.

It is not occasional, that in all listed models the energy of the macromolecule is defined by the use of extrinsic curvatures
of the macromolecule curve. Indeed,  the (linear) density of the energy of
the macromolecule should {\sl a } {\sl priori } depend on the geometric
characteristic of the curves and be independent on the way of their
description. Hence, it should depend on the functions of the curve $\gamma$ which are invariant
under its motions (rotations and translations), i.e. on its extrinsic curvatures ${\kappa}_I$,
 $0<I\leq D-1 $ (where $D$ is the dimensionality of the space) \cite{postnikov}

From this viewpoint the common approach, for taking into account
the external field influence, seems to be ill-defined. It consists
in the naive adding to the energy functional given further below
in (\ref{gactions}) of the potential energy term \cite{yamakawa}
\begin{equation}\label{wlc_ham}
{\cal E}={\cal E}_0 + \int_\gamma  V({\bf r}) ds ,
\end{equation}
where  $V({\bf r(s)})$ is the potential of the external field.

   The geometric (``macroscopic") inconsistency is that
the energy functional of the macromolecule becomes the function of a quantity, which is  not
 invariant under the motions of the curve (rotations and translation in  Euclidean space).

The physical (``microscopic") inconsistency of this approach is the
independence of the chain flexibility (e.g. the persistence length
$l_{p}$) on the external field.
 While the stretching or
the compressing fields substantially restricted the set of available
conformations of each repeated unit of the polymeric chain. Thus,
a strong enough external field should affect the chain
flexibility, and the persistence length $l_{p}$.  This issue
should be relevant, e.g. for the confinement-induced changes of the
persistence length observed in \cite{cifra08}. The effective
compressing field, describing the intra - chain attraction and chain
confinement in restricted geometries yields a substantial
change of the apparent persistence length \cite{kulic,cifra10}.

The purpose of the present paper  is to suggest an alternative way for taking into account the external field influence,
which seems to be free   from the above discrepancies.
Namely, we suggest to involve the external field in the definition of the line element $ds$, associating it with the effective metric
\be\label{effective}
g_{AB}=\left(1+V({\bf r})\right)^2\delta_{AB}\equiv n^2({\bf r})\delta_{AB},
\ee
and to define the  extrinsic curvatures $\kappa_I$, in accordance with Frenet equations,
on the space equipped with this effective metric.

Hence,  we suggest to preserve the functional dependence of the energy density from the
extrinsic curvature, but modify the definition of the
arc-length, and consequently,   of the extrinsic curvatures.

The paper is organized as follows.

In the Second section we
present the explicit expression of the suggested geometric model of the macromolecule interaction with the external
field, and discuss the analogies of this model
with relativistic  particle systems  and with  geometric optics.

In the Third section we give its Hamiltonian formulation  and analyze it in the framework of the Dirac's theory of constrained systems.


\setcounter{equation}{0}
\section{The model}
We suggest to describe the external field influence on the macromolecule by the  energy functional
\begin{equation}
 {\cal E}_0 =\int_\gamma {F}({\kappa}_1,....,{\kappa}_N )d{\widetilde s},
\label{gactions}\end{equation}

where
\begin{itemize}
 \item $F(\kappa_I)$ is the energy density in the initial (noninteracting) DNA model. It can be chosen say,
 as in (\ref{wlc}), (\ref{wlc2}), (\ref{feoli})
 \item
The element of arclength (or control length)   $d{\widetilde s}$ is defined in accordance with the metric (\ref{effective}),
\be
d{\widetilde s}= n({\bf r})|\dot{\bf r}| d\tau\equiv{\dot{\widetilde s}}d\tau,
\ee
where $\tau$ is an arbitrary parameter defining the curve $\gamma$:
${\bf r}={\bf r}(\tau)$

\item The extrinsic curvatures $\kappa_I$ are defined, via Frenet equations, on the space equipped by the effective metric (\ref{effective}),
\begin{equation}
 \frac{{\rm d}{\bf r}}{\dot{ s}{\rm d}\tau}={\bf e}_1,\quad
\frac{{\rm D}{\bf e}_a}{\dot{ s}{\rm d}\tau}={\kappa}_a{\bf e}_{a+1}
-{\kappa}_{a-1}{\bf e}_{a-1}.  \label{ff}\end{equation}
Here
$$
\frac{D}{{\rm d}\tau}\equiv\frac{{\rm d}}{{\rm d}\tau}
+{\widehat\Gamma}({{\bf \dot r}}),\quad ({\widehat \Gamma})^A_B\equiv\Gamma^A_{BC}{\dot x}^C,
 \qquad{\bf e}_a\cdot{\bf e}_b=\delta_{ab},\quad
{\bf e}_0={\bf e}_{D+1}\equiv 0,\ I=1,\ldots, D
$$
with ${\bf e}_i{\bf e}_j\equiv e^{(i)A}g_{AB}e^{(j)B}$, being $D$ the dimensionality of the space and
$\Gamma^A_{BC}$ the Christoffel symbols of the  effective metric
\be
 \Gamma^C_{A B}=n_A
\delta_{CB}+n_B\delta_{CA}-n_C \delta_{AB}, \qquad n_A\equiv
{\partial_A \log n}.
\ee
Similarly to the Euclidean case,
the
curvatures
 $\kappa_1,\ldots, \kappa_{D-1}$   are positive functions, and the highest curvature (torsion),
 $\kappa_{D-1}$ can take both positive and negative values.
If some   $\kappa_I\neq 0$,  then $\kappa_i\neq 0$
$i=1,2,\ldots,I-1$. Vice versa, from  $\kappa_I= 0$,  it follows that $\kappa_i= 0$  $\mu=I+1,\ldots, D-1$.
\end{itemize}

\noindent
Explicit expressions for external curvatures look as follows:
\begin{equation}
{\kappa}_I
=\frac{\sqrt{\det{\hat g}_{I+1}\det{\hat g}_{I-1}}}{\det{\hat g}_I },
 \quad \quad {(g_I)}_{ij}
\equiv
\frac{d^i{\bf r}}{(d{\widetilde s})^i}\frac{d^j{\bf r}}{(d{\widetilde s})^j},\quad
i,j=1,\ldots,I.
\label{ecurve}\end{equation}
In particular,
 the first curvature $\kappa_1$
reads \be \kappa_1=\frac{1}{n^2({\bf
r})}{\sqrt{\kappa^2_{(flat)1}+\frac{{\dot n}^2}{|{\dot{\bf
r}}|^2n^4} }}, \ee where $\kappa_{(flat)1}$ is the first curvature
of the curve on the flat space. Hence, within the suggested approach, we
have  got the geometrically consistent energy functional of the
macromolecule, where   the influence of the external field in the
internal energy is taken into account. Since we deal with the
macroscopic model (or a phenomenological description) of the macromolecule, we
can assume that $n({\bf r})$ parametrically depends on the
temperature $T$. For the correspondence with the existing models we
can choose, e.g. $n=n(T, {\bf r})={k_B T}(1+\frac{V({\bf r})}{k_B
T})$.  The environment influence on the chain elasticity and
flexibility is usually described in terms of the persistence length.
For example, to describe the DNA behavior in nanochannels and nanoslits,
the concept of the global persistence length has been introduced in
\cite{odijkappa08}, as a typical distance between hairpins in the
confined chain. The main advantage of the obtained result is the
deep non-trivial relationship between the bending elasticity and
the external field. The proposed approach seems to be useful for a better
understanding of the mechanism of elasticity of macromolecules, in
stretching and confining fields.\\

The price for the geometrization of the model is the further mathematical complication of the system.
In spite of the apparent simplicity, the analysis of models with the
energy functional (\ref{gactions}) is not a simple task: the matter
is that the energy density $F$ in (\ref{gactions})
 depends on higher  derivatives
and should be analyzed in the framework
 of the Dirac's theory of constrained systems \cite{dirac}.
 Fortunately, in the eighties such functionals were widely studied in
 the context of the (relativistic) spinning particles. Namely, the functionals (\ref{gactions})
 in the Minkowski space-time, with $s$ being a world-line
 parameter were considered. These studies were inspired by the papers
Polyakov devoted to the  Chern-Simons theories
and rigid strings \cite{polyakov}. Various systems  depending on the first and the second
curvatures of the path  in three- and four- dimensional Minkowski spaces  were studied in
details, mostly by Plyushchay (see \cite{misha,mmassless} and references therein).
By this
 reason the relevant formalism  for
analyzing such systems on the base of  the Frenet equations has
been developed by one of the authors  \cite{tmp}. Almost all
systems depending on extrinsic curvatures either by isospin (i.e.
by internal degrees of freedom), or spin (in the case when they
are proportional to a single curvature, see
\cite{mmassless,massless}. In the context of macromolecules it
means, that the systems with energy density depending on extrinsic
curvatures of macromolecules should have internal degrees of
freedom. Now, any macromolecule is a sequence of atoms connected
by chemical bonds. That is why, the universal way of any change of
chain conformation is the change of dihedral angles, caused by
rotation around the chemical bonds. Thus, any polymeric chain has
its internal rotational degrees of freedom, which should be
obviously identified with the isospin of the system. These studies
were performed on flat spaces, mostly. The only exceptions are,
seemingly \cite{curved,massless}. A qualitative observation there
is that, for Lagrangians (in our case: energy densities) with
linear dependence on extrinsic curvatures, the introduction of
non-Euclidean metrics drastically changes the properties of the
system, including the dimensionality of the internal space. Only
for the constant curvature spaces (e.g. spheres, hyperboloids)
non-Euclidean metrics do not change essentially the properties of
such systems. In three-dimensional space, in our notation it
corresponds to the choice \be n({\bf r})=\alpha_0/(1+\beta{\bf
r}^2) \ee corresponding to the three-dimensional sphere or
hyperboloid (depending on the choice of the sign of $\beta $).

The suggested DNA model has an obvious analogy with the geometric optic,
 where  the trajectory of light in the media with
refraction index $n({\bf r})$
corresponds to the minimum of the functional (under assumption that the helicity of light is neglected)
\begin{equation}
 {\cal S}_{Fermat} =\int d{\widetilde s},\qquad d{\widetilde s}=n({\bf r}) |\dot {\bf r}|d\tau .
\label{gactions2}\end{equation} Hence, the generic action
(\ref{gactions}) defined on such ``effectively" curved space, will
take into account the interaction of the spin (helicity) of the
light with the media, i.e. the phenomenon  of the ``optical Hall
effect", consisting in the deviation of the light trajectory from
that given by Fermat Principle, due to the feedback from the
polarization \cite{opt-Hall}. Let us also mention, in this respect,
the recent paper \cite{yaf},
 where the analysis of the model (\ref{feoli}) with a non-constant effective metric
has been performed in some details, in the context of geometric optic.\\

So, we suggest to describe the DNA molecules interacting with the
external field by the functional (\ref{gactions}), where the
arc-length is defined by the expression $d{\widetilde s}\equiv
 (1+V({\bf r} ))|\dot{\bf r}|d\tau$, and the extrinsic
 curvatures $\kappa_1,\ldots,\kappa_N, N<D$  are defined in
 accordance with modified Frenet equations (\ref{ff}).

\setcounter{equation}{0}
 \section{Hamiltonian formulation}
It is seen from the Frenet equations (\ref{ff}), that the $I$-th extrinsic curvature $\kappa_I$ depends on the $(I+1)$-th derivative,
 $\kappa_I=\kappa_I({\bf r}^{(i)}))$, so that
  the energy density $F(\kappa_i)$ in  (\ref{gactions}) depends on the
$(N+1)$-th derivative. By this reason  the study of its Euler-Lagrange equations is not a simple task.
To simplify the analysis of (\ref{gactions}), let us consider the variationally equivalent
functional, whose integrand depends on the first
derivatives only (see \cite{tmp,massless,david} for further details).

Precisely, we replace the initial energy density ${\cal L}$ by the
following  variationally equivalent one:
$$
{\widetilde{F}}=
{{F}}(\kappa_1,\ldots, \kappa_N)-
\lambda\left(\kappa_{N}-\sqrt{\left(\frac{{D{\bf e}}_{N}}{{\dot{\widetilde s}}d\tau}\right)^2 -\kappa^{2}_{N-1}}\right)
$$
\be
+{\bf p}\left(\frac{d{\bf x}}{{\dot{\widetilde s}}d\tau}-{\bf e}_1\right)
+\sum_{i-1}{\bf p}_{i-1}
\left(\frac{D{\bf e}_{i-1}}{{\dot{\widetilde s}}d\tau}-
\kappa_{i-1}{\bf e}_{i}+
\kappa_{i-2}{\bf e}_{i-2}\right)-\sum_{i,j} d^{ij}\left({\bf e}_i{\bf e}_j-\delta_{ij}\right),
\label{lfo}\ee
where  ${\dot{\widetilde s}},  \kappa_{i-1}, d^{ij}, {\bf p}_{i-1}, {\bf e}_i, \lambda $ play the role of independent variables, and
the relation
$$
\kappa^2_{N}=\left(\frac{D{\bf e}_N}{{\dot s }d\tau}\right)^2 -
\kappa^2_{N-1},
$$
which follows from (\ref{ff}) is taken into account.
Here and further below, we use the following notation for the scalar products: ${\bf p}_i\cdot{\bf
p}_j\equiv p_{(i)A}g^{AB}p_{(j)B}$, ${\bf p }_i\cdot{\bf e}_j\equiv
p_{(i)A}e^A_j$, where, upon summation the indices   $A, B$ take the values $1,\ldots,D$, and $i,j=1,\ldots , N$.
\\

It is easy to see, that varying the energy density ${\widetilde F}$ by $\kappa_{N}$, we shall get $\lambda=\partial F/\partial\kappa_N$,
while its variations by  ${\bf p}$,
${\bf p}_{i-1}$ and $d_{ij}$ will yield the Frenet equations restoring the dependence of $\kappa_{i}$ on ${\bf r}^{(i)}$.

The Hamiltonian formulation of the system with the singular ``Lagrangian" (\ref{lfo}) on the Euclidean space was carried out in \cite{tmp}.
Its extension to curved space is straightforward. For the ``Lagrangians" linear on extrinsic curvatures it was considered in \cite{massless},
and for the generic case in \cite{david}.
In accordance with this paper, the Euler-Lagrange equations  of the functional with the energy density
${\widetilde{F}}$ are  given by  the
 constrained Hamiltonian system,
\be
\omega_N=d{\bf p}\wedge d{\bf r}+ \sum_{i=1}^N d{\bf p}_i\wedge
d{\bf e}_i,\qquad {H}={h}_{0}+\sum_{i}\kappa_{i}{\phi}_{i}+ \sum_{i,j} d^{ij}({\bf e}_i{\bf e}_j-\delta_{ij}),
\label{hred}\end{equation}
with the following set of constraints (including some gauge fixing conditions, see for details \cite{tmp,david}):
\be
\quad
{\phi}_{0} \equiv\bpi {\bf e}_1+\sum_{i}{\kappa}_iF_{,i}-F\approx 0,\qquad {\rm where}\;{\bpi}\equiv {\bf p}-{\bf
\Gamma},\quad
 \Gamma_A\equiv \sum\limits_{i=1}^N{\Gamma}^C_{AB}{p}_{(i)C}{e}^B_i\label{phi0}\ee
\be
{\bf p}_i{\bf e}_j\approx 0, \quad i\geq j\label{gauge}\ee
\be
{\bf e}_i{\bf e}_j-\delta_{ij}\approx 0,
\label{ort}\ee
\be
 \phi_{i-1}\equiv {\bf p}_{i-1}{\bf e}_i- {\bf p}_i{\bf e}_{i-1}-F_{,i-1}\approx 0,
 \quad 2\phi_N=\frac{1}{F_{,N}}\left({\bf p}_N{\bf p}_N-\sum_i({\bf p}_N{\bf e}_i)({\bf p}_N {\bf e}_i)\right)-
 F_{,N}\approx 0
\label{sectilde}
\end{equation}
For $\det \partial^2 F/\partial\kappa_i\partial\kappa_j \neq 0$, we can resolve the equations (\ref{sectilde}), expressing
$\kappa_i$ via ${\bf p}_i, {\bf e}_j$. After that we should substitute these expressions for $\kappa_i$ in (\ref{phi0}).

Otherwise, for  ${corank}\;\partial^2 F/\partial\kappa_i\partial\kappa_j\;=M\neq 0 $,  we can resolve only $(N-M)$ equations in
(\ref{sectilde}), while the rest will appear as a primary constraint, with the $M$ undefined  functions on $\kappa_i$
playing the role of Lagrangian multipliers.

After these manipulations we will get the constrained Hamiltonian system with $N(N-1)+1$
primary constraints given by (\ref{phi0})-(\ref{ort})
and with $M$ primary constraints obtained from (\ref{sectilde}).
Then, we should perform  the Dirac's procedure of ``stabilization" of the above set of constraints, i.e. require $\{\ldots, H\}\approx 0$, where
by $\ldots$ we encoded the set of primary constraints. This requirement will either fix the values of Lagrangian multipliers,
 or yield an additional set of ``secondary" constraints. Then, stabilizing the obtained set of secondary constraints, we shall either get the
 new set of secondary constraints, or will further fix the values of Lagrangian multipliers, and so on (see, e.g., \cite{dirac}).

It is easy to observe, that the ``stabilization'' of the constraints (\ref{gauge}) fixes the values of
the Lagrangian multipliers $d^{ij}$, while the stabilization
of the remaining set depends on the concrete properties of the function $F(\kappa_1,\ldots,\kappa_N)$.

For example, for the energy densities with a non-degenerate quadratic dependence on curvatures,
 $F=c_{ij}\kappa_i\kappa_j$, $\det c_{ij}\neq 0$,
which involves, as a particular case, the WLC model (\ref{wlc}) and Buggle-Fujita model (\ref{wlc2}),
 there are no secondary constraints.
The whole set of (primary) constraints is divided in the second-class constraints given by (\ref{gauge}),(\ref{ort}),
and the single first-class constraint  given by (\ref{phi0}).
As a consequence, we get that the dimensionality of the physical phase space is $2D(N+1)-2-N(N-1)=2DN-N(N-1)$.
In contrast with this case, in the systems with  linear energy densities, $F=c_i\kappa_i+c_0$, the whole set of constraints and their
algebra,
essentially depend on the values of the constants $c_i$ and on the properties of the effective metric (\ref{effective}).

Hence,  in contrast with the functionals with linear dependence on curvature,
in the case of quadratic dependence, the introduction of the curved metric does not essentially change the structure of the phase space of the
system, being reflected only on the functional dependence of the Hamiltonian from the external field.

To complete this general consideration, let us present the following useful expressions, which should be useful for the
further analysis of the systems under consideration.

At first, let us   write down the
non-zero
Poisson brackets between the functions ${\bpi e}_i$, ${\bf \pi p}_i$, ${\bf \pi }^2$:
\be
\begin{array}{c}
\{{\bpi }{\bf e}_{i},{\bpi}{\bf e}_{j}\}= {\cal R}({\bf }{\bf
e}_{i}, {\bf }{\bf e}_{j}),\quad \{{\bpi }{\bf p}_{i},{\bpi }{\bf
p}_{j}\}=
{\cal R}({\bf p}_{i}, {\bf p}_{j})\\
\{{\bpi {\bf e}}_{i},{\bpi {\bf p}}_{j}\}= {\cal R}({\bf e}_i, {\bf
p}_j)-\bpi^2\delta_{ij},\quad \{{\bpi }{\bf e}_{i},{\bpi }^2\}= -2
{\cal R}({\bpi}, {\bf e}_i),\quad \{{\bpi {\bf p}}_{i},{\bpi }^2\}=
-2{\cal R}({\bpi}, {\bf p}_i)
\end{array}
\label{pb}\ee
where
\be
 {\cal R}({\bf a},{\bf b})\equiv
\sum_{i=1}^N R({\bf p}_i|{\bf e}_i,{\bf a},{\bf b}).
\label{r}\ee
is the Riemann tensor associated with the effective metric (\ref{effective}).

Then, let us present the expression 
for
the Riemann tensor in the three-dimensional space,
 where it  is uniquely expressed via the Ricci tensor
\begin{equation}
R_{ABCD}=R_{(2)AC}g_{BD}-
R_{(2)AD}g_{BC}+R_{(2)BD}g_{AC}-R_{(2)BC}g_{AD}+\frac{R_0}{2}\left(g_{AD}g_{BC}-g_{AC}g_{BD}\right),
\label{riemann}\end{equation}
where $R_2$  is the Ricci tensor
and $R_0$ is the scalar curvature of the space.
For the conformal flat  metric (\ref{effective}) the  Ricci tensor is  defined by the formulae
\begin{equation}
R_{AC}=
-g_{AC}(n_{DD}+3n_D^{2})-n_{A,C}+n_A n_C,\qquad n_A\equiv
{\partial_A \log n}.
\end{equation}
All considerations above were quite general, with a generic functional
dependence  of the energy density on the extrinsic curvatures,
 and they have been formulated in the $D$-dimensional space.
As an example, let us give  the explicit formulation of the WLC model (\ref{wlc}),
with the additional constant term $c_0$ \be {F}=
\frac{c}{2}\kappa_{1}^2+c_0. \ee
In this case the Hamiltonian system
(\ref{hred}) reads
\be \omega=d{\bf p}\wedge d{\bf r}+ d{\bf
p}_1\wedge d{\bf e}_1,\qquad {H}={\bpi}{\bf e}_1+\frac{{\bf
p}^{2}_1}{2c}-c_0+\left(c_0-\frac{3}{2}{\bpi}{\bf e}_1\right)\left({\bf e}_1\cdot{\bf e}_1-1\right),
\label{}\end{equation}
with the following set of  constraints:
\be {\bf e}_1\cdot{\bf
e}_1-1\approx 0,\quad {\bf p}_1\cdot{\bf e}_1\approx 0,\quad
{\bpi}{\bf e}_1+\frac{{\bf p}^{2}_1}{2c}-c_0\approx 0.\ee
The extrinsic curvatures looks  as follows: \be
\kappa_1=\sqrt{\frac{{\bf p}^{2}_{1}}{c^2}},
\qquad \kappa^{2}_2=\frac{1}{c^2\kappa^{2}_1}\left(   \bpi^2-4c_0\bpi{\bf e}_1+\frac{3}{4}(\bpi{\bf e}_1)^2 +c^2_0\right)-\kappa^2_1.
\ee
The evolution equations for these curvatures could be straightforwardly obtained by the use of Poisson bracket relations (\ref{pb}). They will obviously depend
from the Riemann tensor (\ref{r}).

We suppose to present the detailed study of the WLC model in a forthcoming paper.

\section{Discussion}

The statistical mechanics of the WLC has been developed
by regarding it as a differentiable space curve
of fixed length \cite{gross},\cite{yamakawa1}. If ${\bf r}(s)$ is the radius vector of an
arbitrary point of the curve as a function of the contour
distance $s$ from one end to that point,
persistence length $l_{p}$ can be defined as
a correlation length of the tangent vectors fluctuations \cite{gross}
\begin{equation}
\langle {\bf u}(s){\bf u}(s+r)\rangle\sim e^{-r/l{p}},
\end{equation}
where ${\bf u}(s)$ is the unit vector tangential to the curve at the point $s$.
Thus, $l_{p}$ is statistical quantity strictly dependent on
the set of available chain conformations. It is well known that the intra-chain interactions,
e.g. electrostatic repulsion \cite{seol_PRL04} or geometrical restrictions imposed on the
chain conformation \cite{kulic} renormalise persistent length $l_{p}$. However, these
particular issues addressed before could not provide the systematic approach to the
influence of interactions and external fields on the flexibility and elasticity of
macromolecules.
\par
In our opinion just
the artificial separation of the terms describing chain elasticity and external fields or interactions
presented in (\ref{wlc_ham}) results to the strong necessity of the persistence length
renormalization. Besides of artificial renormalization of the persistence length the commonly used
additive scheme produced numerous inconsistencies with the description of chain flexibility. For example,
stretching experiments with dsDNA of length about $0.6~-~7~\mu m$ produced the
resulting fitted value for the persistence length depended on the contour length of the dsDNA \cite{seol_BJ07}.
The elastic
energy of highly bent dsDNA conformations is lower than predicted by WLC model \cite{wiggins}. The
list of inconsistencies can be continued.

The proposed above approach permits to take into account the
external fields (or interactions) effect on the chain rigidity and
elasticity directly without to invoke the artificial renormalization
of persistence length $l_{p}$ and promised to extend our
understanding of the physical properties of biological
macromolecules. It consists in the introduction of the effective
metrics depending on the potential of the external field, with
further re-definition of the arc-length parameter and of the
extrinsic curvatures of the chain. These redefinitions obviously
can be interpreted as an impact of the external field in the
internal energy of macromolecules, and in the natural changes of the
persistent length. The proposed way to address the effect of the external fields
on the conformational statistics of macromolecules seems to be much more general than
commonly used. Surely, we do not get any concrete prediction on
the physical relevance of the suggested coupling. But we believe in
its correctness, since it was suggested in a first principles in accordance
with quite
general symmetry considerations, and from this point of
view, can not be wrong. While common scheme of counting
external field influence coupling is pure heuristic one, and has no
any  motivation, except formal simplicity (or analogy with
nonrelativistic mechanical systems). Moreover, any other
coupling scheme, based on the ``microscopic consideration", will be in our opinion
{\sl \'a  priory} more rough and simplified, than presented one. The proposed
symmetry -- based approach to the external field influence is in the
same spirit as Bugl and Fujita model \cite{bugl} taking into account
bending and torsion in terms of the first and second curvatures.

Technical attractivity of the suggested DNA model insist in the  with
spinning particles on curved space, as well as with geometric
optics. This  allows us easily to evaluate qualitative properties of
the systems, as well as to involve, in the study of the DNA
macromolecules, the tools of symplectic and Riemann geometry,
Dirac's theory of constrained systems  and of the relativistic quantum mechanics.\\

{\large Acknowledgments}
 We are grateful  to Armen Allahverdyan and Vladimir Morozov  for encouragement and useful comments.
The work was supported by
 and ANSEF-2229PS grant and by
 Volkswagen Foundation  grant I/84~496.

\end{document}